\documentclass[%
reprint,
showpacs,showkeys,
nobibnotes,
amsmath,amssymb,
aip,
apl,
floatfix,
]{revtex4-1}

\usepackage{graphicx}% Include figure files
\usepackage{dcolumn}% Align table columns on decimal point
\usepackage{bm}% bold math
\usepackage{hyperref}% add hypertext capabilities
\usepackage[english]{babel}
\usepackage{blindtext}
\usepackage{diagbox}
\newlength{\picwidth}
\begin{document}
\setlength{\picwidth}{\linewidth}
\preprint{APS/123-QED}

\title{Film thickness of Pb islands on the Si(111) surface}

\author{Th. Sp\"ath}
\affiliation{Physikalisches Institut, Karlsruhe Institute of Technology (KIT), 76131 Karlsruhe, Germany}
\affiliation{Present address: Department of Materials Science, Darmstadt University of
Technology, Jovanka-Bontschits-Str. 2, 64287 Darmstadt, Germany}

\author{M. Popp}
\affiliation{Physikalisches Institut, Karlsruhe Institute of Technology (KIT), 76131 Karlsruhe, Germany}
\affiliation{Present address: Department of Physics,
University of Erlangen-Nuremberg,
Staudtstr. 7, 91058 Erlangen, Germany}

\author{R. Hoffmann-Vogel}
\email{hoffmannvogel@uni-potsdam.de}
%\affiliation{Physikalisches Institut, Karlsruhe Institute of Technology (KIT), 76131 Karlsruhe, Germany}
\affiliation{Department of Physics, University of Konstanz, Universit\"atsstrasse 10,
78464 Konstanz, Germany}
\affiliation{Institut f\"ur Physik und Astronomie, Universit\"at Potsdam, Karl-Liebknecht-Str. 24-25, 14476 Potsdam, Germany}

\date{\today}

\begin{abstract}
We analyze topographic scanning force microscopy images together with Kelvin probe images obtained on Pb islands and on the wetting layer on Si(111) for variable annealing times. Within the wetting layer we observe negatively charged Si-rich areas. We show evidence that these Si-rich areas result from islands that have disappeared by coarsening. We argue that the islands are located on Si-rich areas inside the wetting layer such that the Pb/Si interface of the islands is in line with the top of the wetting layer rather than with its interface to the substrate. We propose that the Pb island heights are one atomic layer smaller than previously believed. For the quantum size effect bilayer oscillations of the work function observed in this system, we conclude that for film thicknesses below 9 atomic layers large values of the work function correspond to even numbers of monolayers instead of odd ones. The atomically precise island height is important to understand ultrafast "explosive" island growth in this system.
\end{abstract}

%\keywords{Scanning Force Microscopy}
\maketitle

%Introduction

\begin{figure}
\includegraphics[width=0.8\linewidth,angle=0,clip]{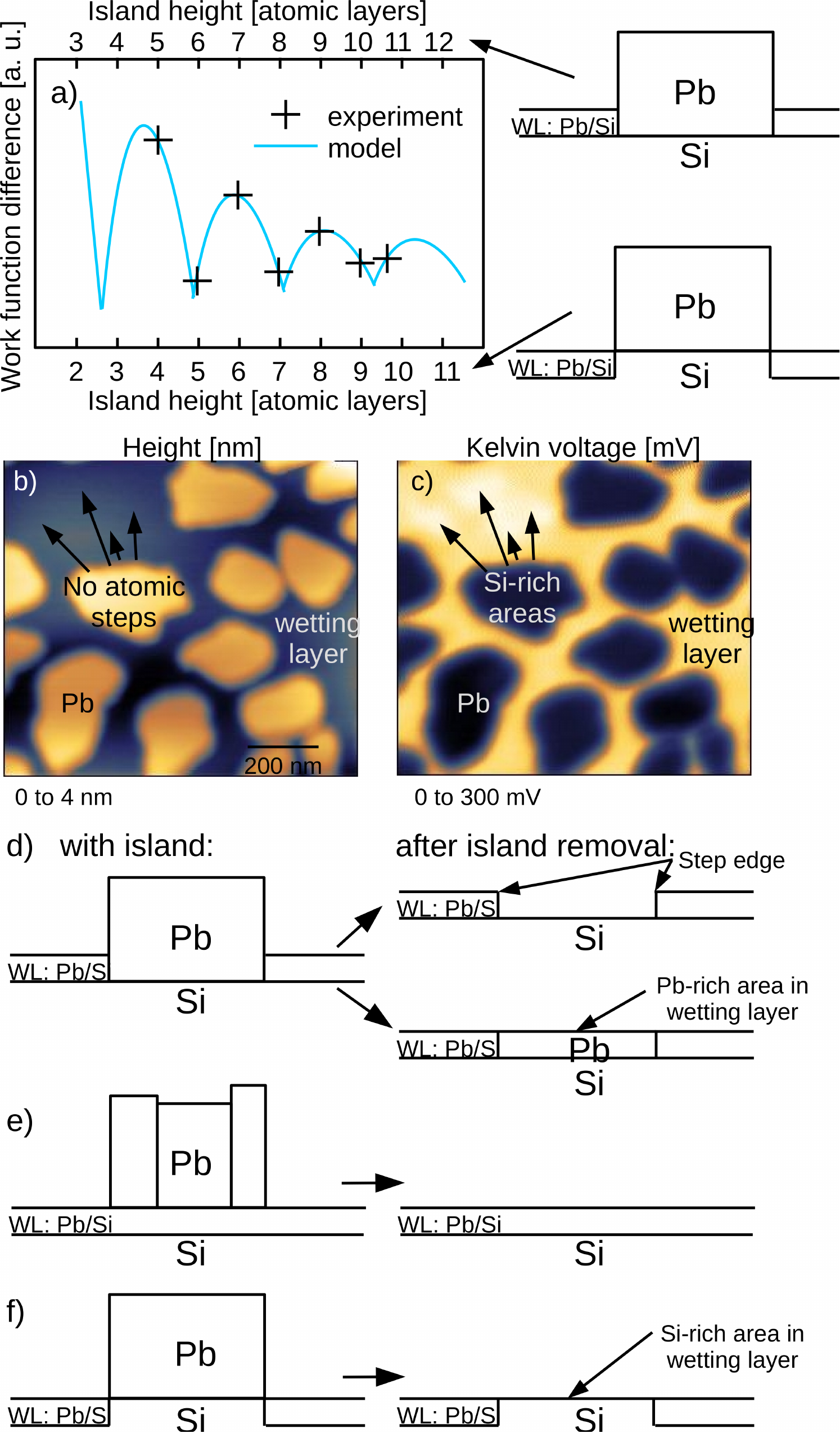}
\caption{a) The quantum size effect leads to a bilayer oscillation of the work function with atomic-scale film thickness in Pb on Si. Top and bottom horizontal axes represent two different models leading to two different phases of the oscillation. b) SFM topographic image obtained after long annealing time, $6$ h $50$ min, ($F= 50$ nA) and c) corresponding KPFM image. $f_0=283$~kHz, $\Delta f=-37$~Hz, $A= 15$~nm. d) Growth of Pb on Si(111) where the island is inserted in the wetting layer. e) Growth of Pb on the wetting layer. f) Growth of Pb islands on Si-rich areas inside the wetting layer. This scheme is supported by our measurements.}
\label{fig_1}
\end{figure}

The Pb on Si(111) system shows the quantum size effect in the local work function \cite{jalochowski88p1,spaeth17p1}.  Due to the confinement of the electrons between the Pb/Si substrate interface and vacuum, the local work function shows a bilayer oscillation, Fig. \ref{fig_1}a). In scanning probe experiments the island height is measured from the top of the wetting layer. Since the position of the island's Pb/Si interface near the substrate is not well-known, the phase of the bilayer oscillation of the work function has been debated, see Fig. \ref{fig_1}a). The growth of Pb on Si is of general interest due to the covalent nature of Si-Si bonds versus the metallic nature of Pb-Pb bonds. In comparison, for Si$_x$Ge$_{1-x}$, for instance, surface alloying has been found to be the driving force for lateral motion of islands \cite{denker05p1}. Shape changes related to interdiffusion and strain changes have been investigated in thermodynamic and kinetic models \cite{magalhaes02p1}. 

The Pb on Si(111) system shows additional peculiarities: The wetting layer is approximately one atomic layer thick, while it contains 1.2 atomic layers of pure Pb\cite{hattab16p1}. In mixed PbSi surface phases the Pb content can be larger than in pure Pb\cite{tong99p1}. For growth on the Si(111)$7\times 7$ reconstruction, the reconstruction remains intact after Pb deposition\cite{hupalo05p1}. The quantum size effect in Pb/Si influences island growth and coarsening \cite{hupalo01p1,chang02p1,jeffrey06p1}. In addition, collective superdiffusion involving ultrafast spreading or a liquid-like diffusion has been discussed in the Pb/Si system \cite{man13p1,huang12p1}. As additional Pb is added by deposition during growth, the islands grow unexpectedly fast ("explosive nucleation") \cite{hershberger14p1}. Depending on the atomically precise island height, the origin of the additional Pb added during growth is interpreted differently.

The atomically precise island height has been investigated previously using X-ray scattering \cite{feng04p1}. It was found that the islands grow on the Si substrate directly and not on the wetting layer \cite{hattab16p1}. In addition, the cristallinity of the islands was confirmed. Whether the islands grow on Si-rich areas inside the wetting layer or within the wetting layer has remained an open question. In scanning tunneling microscopy (STM) measurements \cite{su01p1} the apparent tunneling decay length depends on the tunneling voltage due to electronic effects \cite{kim10p1}. Since the work function is obtained from the tunneling decay length, it becomes more difficult to find out whether even or odd numbers of atomic layers correspond to large work functions \cite{kim10p1}. In addition, models provide contradictory results \cite{zhang05p1,czoschke05p1}.

Here, we focus on the question of whether the interface of the Pb islands with the Si surface is aligned with the top of the wetting layer or with the wetting layer's interface with the Si substrate. We show scanning force microscopy (SFM) together with Kelvin probe force microscopy (KPFM) measurements of Pb islands and of the wetting layer on Si(111). KPFM allows us to determine the local work function with respect to a reference. The local work function is element-specific and we use it to determine Si-rich areas. We observe Si-rich areas within the wetting layer with the same topographic height as the surrounding wetting layer. We argue that these areas could result from the disappearance of Pb islands located on top of Si-rich areas within the wetting layer. These arguments point to an alignment of the Pb/Si interface with the top of the wetting layer rather than with the wetting layer/Si interface.

%Experimental

We cleaned n-type Si(111) crystals (n-doped using P, $ \rho = 7.5~\Omega$cm) by briefly heating to $1200^{\circ}$C in our vacuum chamber with base pressure of around $3\times10^{-8}$ Pa such that the $7\times 7$-reconstruction was formed as confirmed by SFM. Then one group of the samples were cooled down to room temperature, while a second group were cooled to liquid nitrogen temperatures (LN). The samples cooled to LN were subsequently warmed up in a room temperature environment for 20 to 40 min for annealing. For some exceptional experiments we extended the warming up time to up to 14 h 30~min. We then deposited Pb using electron beam evaporation (Oxford Applied Research, equipped with ion flux monitor). The Pb flux $F$ was varied between 10 and 75~nA. Immediately after deposition the samples were transferred to the SFM (Omicron VT-AFM) attached to the same vacuum vessel. The SFM had been pre-cooled using liquid nitrogen and the measurements were performed at 115~K. The annealing times given in the following include all Pb deposition and transfer times.

We used commercial Si cantilevers (Nanosensors) with a longitudinal spring constant of about 50 N/m and a resonance frequency of approximately 300 kHz. Most of the tips were coated with a 25 nm thick Pt/Ir layer to enhance electrical conductivity. The tips were cleaned in vacuum by heating to $150^{\circ}$C for several
hours and by Ar ion sputtering. For SFM imaging we used dynamic frequency modulation mode, oscillating the tip at resonance and keeping the oscillation amplitude constant at 5-15~nm. A Nanonis phase-locked loop (SPECS, Zurich) was used to detect the frequency shift induced by the tip-sample interaction. We performed KPFM measurements simultaneously with the topography measurements with a voltage oscillating at a frequency of $f_{\mbox{\scriptsize Kelvin}} = 266$~Hz and an amplitude of 0.5 V applied to the tip. The sign of the measured Kelvin voltage was known because we have applied a voltage to the tip. The polarity of the final result has been checked previously \cite{perez16p1}. By introducing a negative sign to the data, we represent negative charge by bright contrast, i.e. higher Kelvin voltage.

%Results

Fig. \ref{fig_1}b) shows a topographic image together with the simultaneously measured KPFM image c) obtained after long annealing times (6 h 50 min). In Fig. \ref{fig_1}c) Pb islands appear black showing that Pb is positively charged compared to the PbSi mixed wetting layer (yellow). Within a yellow region, at the position where one might expect a Pb island, there are some noticeably fainter regions of a smaller size than Pb islands. We argue that these fainter regions were freed from Pb islands during coarsening. Possibly, the Pb of these regions has joined other islands nearby or exchanged with the wetting layer.

The fainter regions point to Si-rich areas. The work function of Si (4.85 eV for n-doping) is larger than that of Pb (4.25 eV) \cite{crc08p1}. Indeed the wetting layer shows a larger Kelvin voltage compared to the Pb islands and the Si-rich areas show an even larger Kelvin voltage. The position of the Fermi level is mainly governed by the Pb content \cite{tong99p1,choi07p1}. In our experimental results, the work function within the wetting layer varies by only a few meV. For the small work function differences we obtain here within the wetting layer, we expect that the fainter regions are indeed more Si-rich. Our measurement method is most sensitive to the topmost layer, but electrostatics of lower layers could additionally influence the measurement. Due to the large concentration of Pb in the topmost layer, the wetting layer, and due to the electronic reflection at the Pb/Si interface relevant for the quantum size effect, concentration changes in the wetting layer seem to be most relevant.

\begin{figure}[t]
\includegraphics[width=0.8\linewidth,angle=0,clip]{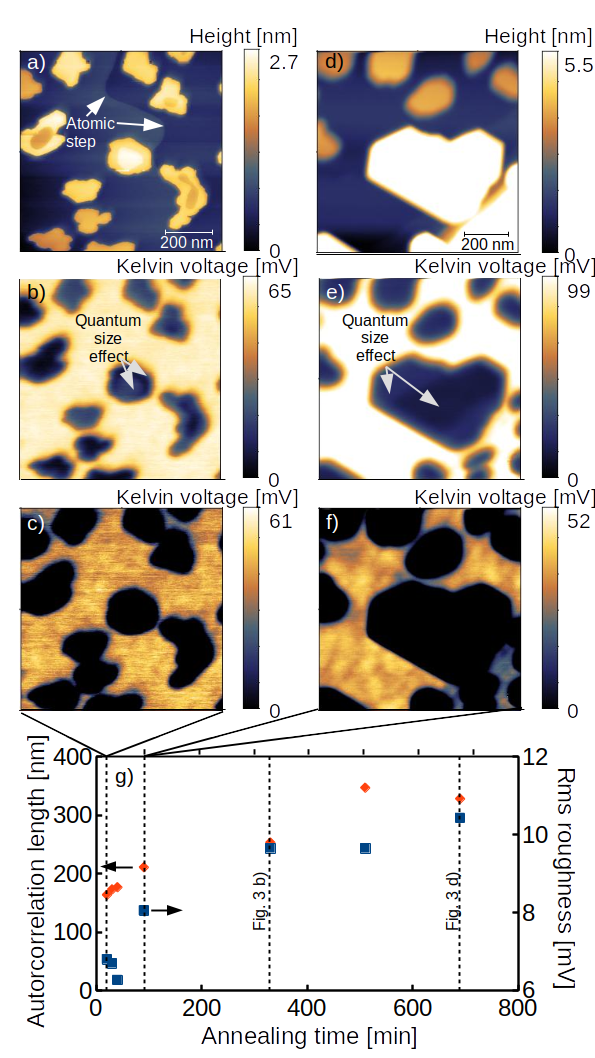}
\caption{a) and d) SFM topographic image obtained after warming up for 20 min ($F=27$ nA) and for 1 h 30 min ($F= 50$ nA). b), c), e) and f) corresponding Kelvin microscopy image using two different color scales. Images b) and e) show quantum size effects on the islands. Images c) and f) focus on the wetting layer. For a) $f_0=280$ kHz, $\Delta f=-20$~Hz, $A= 10$~nm, %$\gamma=-3.6$ fNm$^{(1/2)}$,
$c_L= 50$ N/m. For d) $f_0=283$~kHz, $\Delta f=-27$~Hz, $A= 15$~nm. g) The root mean square roughness (blue squares) and the autocorrelation length (red rhombs) plotted as a function of annealing time. Both the root mean square roughness and the autocorrelation length increase as a function of annealing time.}
\label{fig_2}
\end{figure}

\begin{figure}
\includegraphics[width=0.8\linewidth,angle=0,clip]{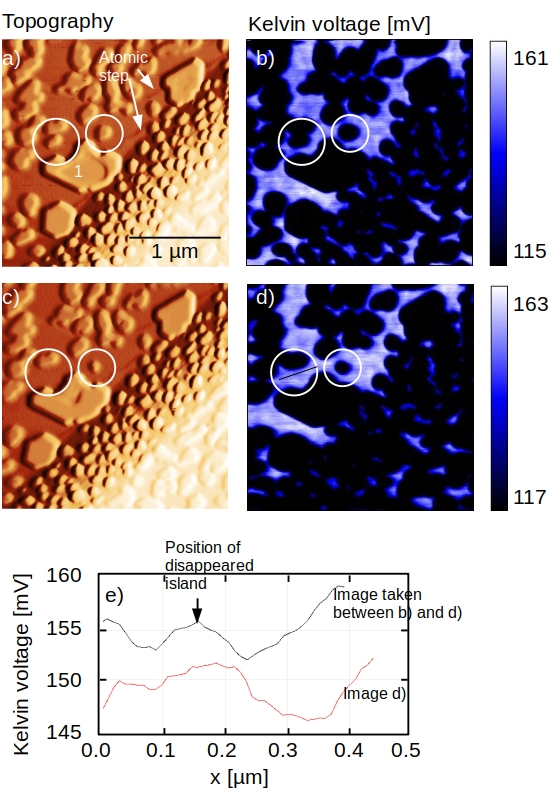}
\caption{a) SFM topographic image obtained during warming up the sample starting from 115 K after 5 h 30 min ($F= 50$ nA) b) Corresponding KPFM image where the color scale was chosen to focus on the wetting layer. Two islands have been marked by circles in a) and b). An island marked with 1 has an outer rim of greater height. c) A subsequent topographic SFM image to a) acquired after 11~h 30 min and d) the corresponding KPFM image. The same positions as in a) and b) have been marked by circles. One island has decayed, another has become significantly smaller. At both positions a larger Kelvin voltage is observed. A shadow has been added artificially to a) and c) to highlight atomic steps. Acquisition time 3 h. e) Line profile extracted at the position of the black line in part d) and a line profile taken at the same position from the image acquired between b) and d). The line profiles show a local maximum of contrast at the position where the island has disappeared. $f_0=283$~kHz, $A=15$~nm a) $\Delta f=-33$~Hz, b) $\Delta f=-34$~Hz.}
\label{fig_3}
\end{figure}

While we routinely observe atomic steps on the wetting layer, see Fig. \ref{fig_2}a and \ref{fig_3}a, the surface shows no atomic steps at the position of the high Kelvin voltage regions (Fig. \ref{fig_1}b)). Fig. \ref{fig_1}d) to f) show different possible scenarios for the arrangement of Pb islands on Si(111). The arrangement shown in Fig. \ref{fig_1}d) has been favored in literature \cite{feng04p1}. In this scenario we would expect to observe either an atomic step edge or a Pb-rich area after removal of an island. The scenario shown in Fig. \ref{fig_1}e) has been ruled out previously \cite{feng04p1}. In this scenario the interface of the Pb islands to the wetting layer substrate would be rough. This would contradict a variety of experimental observations. The presence of the quantum size effect requires excellent reflection at the top and bottom surface of the island. In Low Energy Electron Diffraction experiments, the islands are observed to be crystalline \cite{hattab16p1}. In our {SFM images} we detect polycrystalline islands from subatomic steps on the islands' top and the overall shape of apparently merged islands \cite{spaeth17p1}. Most islands do not show such steps and shapes.

The scenario shown in Fig. \ref{fig_1}f) has also been discussed previously \cite{feng04p1}. Only in scenario f), after disappearance of an island, a Si-rich area could be observed if we assume that Pb diffusion did not completely fill the area at the time of observation. Since we neither observe island-shaped atomic steps nor Pb-rich areas within the wetting layer at positions where islands have disappeared or are expected but some Si-rich areas, we argue that the islands are located on Si-rich areas inside the wetting layer (Fig. \ref{fig_1}f)).

We have acquired images systematically as a function of the annealing time (Fig.~\ref{fig_2}). On the islands, influences of the quantum size effects are observed due to atomic steps at the top or the bottom of an island (Fig. \ref{fig_2} b) and e)), for details, see \cite{spaeth17p1}. In KPFM images, Fig. \ref{fig_2} b) and e), the color scale has been arranged in order to make these quantum size effects visible. In Fig. \ref{fig_2} c) and f), the color scale of the same measurements has been arranged to focus on the variations within the wetting layer.

We first applied a threshold mask to each Kelvin image to separate wetting layer and  Pb islands using gwyddion. We then calculated the root mean square roughness and the autocorrelation length only on the wetting layer, see Fig. \ref{fig_2}g). The autocorrelation length is measured at a threshold that was set to the same value for all images. While the first three data points represent an average taken from 3-4 images each, the last four data points represent the images shown in Fig. \ref{fig_2}f), Fig. \ref{fig_3}b), another consecutive image and Fig. \ref{fig_3}d). Both the root mean square roughness and the autocorrelation length increase as a function of annealing time. This suggests that the size of the correlated areas increases as a function of time. As a function of time, island coarsening occurs, i.e. large islands grow at the expense of small islands that ultimately disappear and the size of the islands which have disappeared grows. We believe that the size of the correlated areas and the size of the islands that have disappeared are related. The autocorrelation function in the time domain has been used to study coarsening in this system \cite{tringides10p1}. The results confirm diffusive coarsening at temperatures at and above room temperature.

We have studied the disappearance of an island during imaging while the sample was slowly warmed up starting from a temperature of 115~K. The amount of changes between two consecutive images can be adjusted by chosing an appropriate annealing temperature and duration. Fig. \ref{fig_3}a) shows an area of the Pb-covered Si surface showing islands. A subsequent image is shown in Fig. \ref{fig_3}c). In this image, an island marked by a circle has disappeared, and another one has become significantly smaller. A larger island nearby (marked by 1 in Figure \ref{fig_3}a) has an outer rim of greater height compared to its inner area. The outer rim is by five atomic layers of Pb higher than the inner part of the island as measured in line-sections. The inner area of reduced height has become smaller in size in \ref{fig_3}c) compared to \ref{fig_3}a). At the positions marked by circles in the corresponding Kelvin voltage images, \ref{fig_3}b) and d), we investigated whether the wetting layer shows a larger Kelvin voltage near the area freed by the removal of Pb island material. We show a line profile of the Kelvin voltage measured in this area in Fig. \ref{fig_3}e) taken at the position of the black line in e) and from a consecutive image acquired between part b) and e). Both line profiles clearly show a local maximum at the position where the island has disappeared.

For the Si$_x$Ge$_{1-x}$ system, Si-rich areas located under the metallic islands have also been observed in literature \cite{denker05p1}. In this system, Ge was removed by selective etching. By ex-situ SFM measurements, Si-rich plateaus were observed at the position where islands have disappeared.

We use a situation taken from Ref. \onlinecite{hershberger14p1} as an example: We consider an island with 4 atomic layers thickness and 70 nm diameter that nucleated from the one atomic layer-thick wetting layer with a Pb content of 1.22 atomic layers. The Pb content of all islands was 2.2 times larger than the total amount of Pb deposited. If all of the Pb content of the surrounding wetting layer were used for island growth, Pb would result from a ring with outer diameter of 94 nm (model in Fig. \ref{fig_1}f)) or 117~nm for the model in Fig. \ref{fig_1}d) respectively. This means that in the first case Pb would travel over a distance of 12 nm at most compared to 23.5~nm in the second case. The "explosion" would in the first model be mainly caused by an expulsion of Pb from the wetting layer and the travelling distances of Pb are significantly smaller.

We conclude that the wetting layer shows Si-rich negatively charged areas. The structure depicted in Fig. \ref{fig_1}f) could best represent the experimental situation. With this model of the surface, the number of layers in quantum size effect measurements becomes identical to the number of layers counted from the top of the wetting layer. Simple models of the quantum size effect are in agreement with experimentally observed physics - no additional phase jumps are required. In addition, this model implies an additional reservoir for Pb during growth since the Pb is expelled from one atomic layer as the islands grow. This adds missing Pb to scaling theories, helps to account for extremely fast diffusion and has an impact on the discussion on superdiffusion in this system. 
 
%acknowledgements
We thank C. P\'{e}rez Le\'{o}n for initiating the idea for this analysis, and C. P\'{e}rez Le\'{o}n and M. Marz for enlightening discussions and for help with the experimental work. This work was supported by the ERC Starting Grant NANOCONTACTS (No. 239838) and by the German Science Foundation (DFG) through a Heisenberg fellowship.


\begin{thebibliography}{99}

\bibitem{jalochowski88p1}
M. Ja\l{}ochowski and E. Bauer, Phys. Rev. B {\bf 38}, 5272 (1988)

\bibitem{spaeth17p1}
Th. Sp\"ath, M. Popp, C. P\'erez Le\'on, M. Marz, and R. Hoffmann-Vogel, Nanoscale {\bf 9}, 7868 (2017)

\bibitem{hupalo01p1}
M. Hupalo, S. Kremmer, V. Yeh, L. Berbil-Bautista, E. Abram, and M. C. Tringides, Surf. Sci. {\bf 493}, 526 (2001)

\bibitem{chang02p1}
S. H. Chang, W. B. Su, W. B. Jian, C. S. Chang, L. J. Chen, and T. T. Tsong, Phys. Rev. B {\bf 65}, 245401 (2002)

\bibitem{jeffrey06p1}
C. A. Jeffrey, E. H. Conrad, R. Feng, M. Hupalo, C. Kim, P. J. Ryan, P. F. Miceli, and M. C. Tringides, Phys. Rev. Lett. {\bf 96}, 106105 (2006)

\bibitem{man13p1}
K. L. Man, M. C. Tringides, M. M. T. Loy, and M. S. Altman, Phys. Rev. Lett. {\bf 110}, 036104 (2013)

\bibitem{huang12p1}
L. Huang, C. Z. Wang, M. Z. Li, and K. M. Ho, Phys. Rev. Lett. {\bf 108}, 026101 (2012)

\bibitem{hershberger14p1}
M. T. Hershberger, M. Hupalo, P. A. Thiel, C. Z. Wang, K. M. Ho, and M. C. Tringides, Phys. Rev. Lett. {\bf 113}, 236101 (2014)

\bibitem{denker05p1}
U. Denker, A. Rastelli, M. Stoffel, J. Tersoff, G. Katsaros, G. Costantini, K. Kern, N. Y. Jin-Phillip,
D. E. Jesson, and O. G. Schmidt, Phys. Rev. Lett. {\bf 94}, 216103 (2005)

\bibitem{magalhaes02p1}
R. Magalh\~aes-Paniago, G. Medeiros-Ribeiro, A. Malachias, S. Kycia, T. I. Kamins, and R. S. Williams, Phys. Rev. B {\bf 66}, 245312 (2002)

\bibitem{hattab16p1}
H. Hattab, M. Hupalo, M. T. Hershberger, M. Horn von Hoegen, and M. C. Tringides, Surf. Sci. {\bf 646}, 50 (2016)

\bibitem{tong99p1}
X. Tong, K. Horikoshi, and Sh. Hasegawa, Phys. Rev. B {\bf 60}, 5653 (1999)

\bibitem{hupalo05p1}
M. Hupalo, V. Yeh, T. L. Chan, C. Z. Wang, K. M. Ho, and M. C. Tringides, Phys. Rev. B {\bf 71}, 193408 (2005)

\bibitem{feng04p1}
R. Feng, E. H. Conrad, M. C. Tringides, C. Kim, and P. F. Miceli, Appl. Phys. Lett. {\bf 85}, 3866 (2004)

\bibitem{su01p1}
W. B. Su, S. H. Chang, W. B. Jian, C. S. Chang, L. J. Chen, and Tien T. Tsong, Phys. Rev. Lett. {\bf 86}, 5116 (2001)

\bibitem{kim10p1}
J. Kim, S. Qin, W. Yao, Q. Niu, M. Y. Chou and C. K. Shih, Proc. Natl. Acad. Sci. U. S. A. {\bf 107}, 12761 (2010)

\bibitem{zhang05p1}
Y.-F. Zhang, J.-F. Jia, T.-Zh. Han, Zh. Tang, Q.-T. Shen, Y. Guo, Z. Q. Qiu, and Q.-K. Xue, Phys. Rev. Lett. {\bf 95}, 096802 (2005)

\bibitem{czoschke05p1}
P. Czoschke, H. Hong, L. Basile, and T.-C. Chiang, Phys. Rev. B {\bf 72}, 075402 (2005)

\bibitem{perez16p1}
C. P\'erez Le\'on, H. Drees, S. M. Wippermann, M. Marz, and R. Hoffmann-Vogel, J. Phys. Chem. Lett. {\bf 7}, 426 (2016)

\bibitem{crc08p1}
CRC Handbook of Chemistry and Physics 2008, p. 12–114

\bibitem{choi07p1}
W. H. Choi, H. Koh, E. Rotenberg, and H. W. Yeom, Phys. Rev. B {\bf 75}, 075329 (2007)

\bibitem{tringides10p1}
M. C. Tringides, and M. Hupalo, J. Phys.: Condens. Matter {\bf 22}, 264002 (2010)

\end{thebibliography}
\end{document}